\documentclass[aps,prb,amsmath,amssymb,superscriptaddress]{revtex4}
\usepackage{epsf}
\usepackage{epsfig}
\usepackage{graphicx}

\begin{document}


\title{AC transport in carbon-based devices: challenges and perspectives}


\author{Luis E. F. Foa Torres}
\affiliation{Institute for Materials Science and Max Bergmann Center of Biomaterials, Dresden University of Technology, D-01062 Dresden, Germany}
\author{Gianaurelio Cuniberti}
\affiliation{Institute for Materials Science and Max Bergmann Center of Biomaterials, Dresden University of Technology, D-01062 Dresden, Germany}


\date{\today}

\begin{abstract}
\noindent{\bf Abstract}
\vskip 0.5\baselineskip
Time-dependent fields are a valuable tool to control fundamental quantum
phenomena in highly coherent low dimensional electron systems. Carbon nanotubes and graphene are a promising ground for these studies. Here we offer a brief overview of driven electronic transport in carbon-based materials with the main focus on carbon nanotubes. Recent results predicting control of the current and noise in nanotube based Fabry-P\'{e}rot devices are highlighted.
\vskip 0.5\baselineskip
\noindent{\bf R\'esum\'e}
\vskip 0.5\baselineskip
\noindent
Les champs d\'{e}pendant du temps sont des outils pr\'{e}cieux pour le contr\^{o}le des ph\'{e}nom\`{e}nes quantiques dans les syst\`{e}mes d'\'{e}lectrons fortement coh\'{e}rents \`{e} base dimentionalit\'{e}e. Les nanotubes de carbone et le graph\`{e}ne constituent des objets de base prometteurs pour ces \'{e}tudes. Ici, nous proposons une br\`{e}ve synth\`{e}se des ph\'{e}nom\`{e}nes de transport \'{e}lectronique induits dans les mat\'{e}riaux carbon\'{e}s, en se focalisant sur les nanotubes de carbone. Nous mettons en avant des r\'{e}sultats r\'{e}cents pr\'{e}disant le contr\^{o}le du courant et du bruit dans les r\'{e}sonateurs Fabry-P\'{e}rot \`{a} base de nanotubes.

\end{abstract}


\maketitle

\section{Introduction}

During recent decades the study of electronic transport has flourished in the
arena of low dimensional systems. This was mainly due to the advent of
miniaturization techniques that allowed the exploration of transport
properties of semiconductor heterostructures and quantum dots, and single
molecules contacted to electrodes. Most of the experimental and theoretical
studies were focused on time-independent transport. In contrast, much less
attention has been devoted to the influence of time-dependent excitations by
electromagnetic fields or gate voltages. Despite that, interest has been
steadily growing \cite{Platero2004,Kohler2005} and many captivating phenomena
such as photon-assisted tunnelling, coherent destruction of tunnelling
\cite{Grossmann1991} and quantum charge pumping
\cite{Thouless1983,Altshuler1999,Switkes1999} have been unveiled.
Time-dependent excitations provide an opportunity to achieve control through
selective excitations, opening an avenue for both fundamental research and
practical applications. From the point of view of fundamental research, the
study of the interplay between external fields and the transport mechanisms
offer a rich prospective to improve the understanding of electron dynamics at
the molecular scale. On the other hand, the use of these external fields as
means to control currents in coherent conductors may offer innovative
alternatives for practical applications such as switches \cite{delValle2007},
interconnects \cite{Coiffic2007} and sensors \cite{Kibis2007}.

The feasibility of the implementation for a given material being the ultimate
limitation to the actual realization of such innovations, it is therefore
important to carefully select the physical objects for study. Carbon based
materials such as polymers, graphitic structures and fullerenes are promising
materials. Within the family of fullerenes, single-walled carbon nanotubes
\cite{Charlier2007,SaitoBook1998}, hollow cylinders of nanometer scale
diameter formed by rolling a graphene sheet (nanotubes hereafter), stand as
one of the major driving forces of both fundamental and applied research due
to their outstanding mechanical and electrical properties
\cite{Charlier2007,SaitoBook1998}. Besides nanotubes, graphene is also
emerging as a preferred choice due to its peculiar electronic structure and
the better possibilities for building reproducible contacts to electrodes
\cite{Geim2007,Katsnelson2007,Cresti2008} and could reveal remarkable
phenomena when exposed to AC fields.

In this paper we focus on the case of ac transport through single walled
carbon nanotubes, being most of the effects that we show generalizable to
other classes of low-dimensional systems. First, we point out some of the
novel phenomena unique to AC\ fields and then mention the main challenges as
well as the tools for the description of driven transport in carbon-based
devices. This adresses the questions: why AC fields and how. In Sec. 3 we
offer a brief review of our recent research in this promising area giving a
concrete example for the case of Fabry-P\'{e}rot devices \cite{FoaTorres2009}.

\subsection{Novel phenomena under AC\ fields}

In the context of quantum transport, time-dependent phenomena can be
classified according to their origin \cite{Buettiker2000}: they can arise as a
result of an external driving (applied voltage, time-dependent magnetic flux,
etc.) or may occur spontaneously (frequency dependent shot noise
\cite{Blanter2000}). Our main interest is on the former situation.

One of the most studied phenomena produced by external driving is
\textit{photon-assisted tunnelling} (PAT). The first experiments showing PAT
were carried out during the 1960s in superconductor-insulator-superconductor
structures \cite{Dayem1962}. Motivated by these experiments, Tien and Gordon
\cite{TienGordon} proposed a simple theory for PAT, the main idea is that a
time-dependent potential can induce inelastic tunnelling events by allowing
the exchange of energy quanta (photons) between electrons and the oscillating
field. Since then, photon-assisted tunnelling has proven to be ubiquitous in
electronic transport and it has been observed in double barrier devices,
molecules and, more recently, in carbon nanotubes \cite{Meyer2007}.

Another captivating effect is \textit{quantum charge pumping}
\cite{Thouless1983,Altshuler1999}. A direct current (dc) is usually associated
to a dissipative flow of the electrons in response to an applied bias voltage.
However, in systems of mesoscopic scale a dc current can be generated even at
zero bias. This intriguing quantum coherent effect is called quantum pumping
and a device capable of providing such effect is termed a quantum pump.
Quantum pumping has attracted much attention
\cite{Brouwer1998,Levinson2000,Moskalets2002,Martinez-Mares2004,Kashcheyevs2004,Strass2005,Arrachea2007a}
and experiments aimed at observing this phenomenon have been carried out in
semiconductor quantum dots \cite{Switkes1999}, quantum wires
\cite{Blumenthal2007,Kaestner2008} and also in carbon nanotubes
\cite{Leek2005}. Quantum spin pumping is also a very active area of research
\cite{Blaauboer2005,Das2006,Romeo2008}.

The operational regime of a pump can be characterized according to the
relative magnitude between the driving frequency $\Omega$ and the inverse of
the traversal time through the sample, $1/\tau_{T}$. When $\Omega\ll1/\tau
_{T}$ , the pump is in the so-called adiabatic regime, whereas the opposite
case, $\Omega\gg1/\tau_{T}$, the pump is in the nonadiabatic regime. For a
cyclic adiabatic change of the conductor parameters, the charge pumped per
cycle is determined by the area enclosed in parameter space \cite{Brouwer1998}%
. Beyond the adiabatic regime, pumping has been also studied both
theoretically
\cite{Falko1989,Levinson2000,Vavilov2001,Strass2005,FoaTorres2005,Moldoveanu2007,Agarwal2007}
and experimentally \cite{vanderWiel1999,Kaestner2008}. Although currents
obtained in the nonadiabatic regime are naturally higher, keeping a low
current noise becomes crucial \cite{Strass2005} to obtain useful devices.

Finally, another issue of interest that we would like to point out is the
\textit{phase sensitivity of the current noise}. While for the case of a
static conductor the noise depends only on the transmission probability and
not on the phase, for a driven conductor the current noise is phase sensitive
\cite{Kohler2005}. Later on we will show an example of our recent research
highlighting this issue.

\subsection{Challenges in the description of driven transport through
nanotubes and beyond}

Transport through carbon nanotubes in the presence of time-dependent
excitations remains much less explored than its time-independent counterpart.
Experimental studies include the response to microwave fields
\cite{Kim2004,Yu2005,Meyer2007} and quantum pumping in response to surface
acoustic waves \cite{Leek2005}. Although some theoretical studies addressing
these topics are available in the literature
\cite{Roland2000,Orellana2007,Wuerstle2007,Guigou2007,Oroszlany2009}, the present understanding is not complete. Besides, as we will see later, many of the results obtained for carbon nanotubes could apply to the case of driven
transport in graphene devices, a topic of much current interest
\cite{Trauzettel2007,Lopez-Rodriguez2008,Shafranjuk2008}.

The main challenges for a proper theoretical description stem from the fact
that, besides the need of having an adequate modelling of the electronic
structure, the most interesting regime lies beyond the scope of either low
driving frequencies (adiabatic limit) or perturbative approaches. Although
ambitious, any theoretical advance in this direction will be useful not only
in the context of carbon nanotubes, but in the broader realm of molecular
electronics \cite{Cuniberti2005} and therefore we decided to address this problem.

\subsection{Fabry-P\'{e}rot regime in carbon nanotubes and the influence of AC
fields}

Low resistance contacts are very difficult to achieve in molecular systems.
Indeed, most of the experiments in molecular electronics are in the
Coulomb-Blockade regime where coupling between the molecule and the electrodes
is weak. However, carbon nanotubes offer a unique opportunity to
experimentally explore high quality low resistance contacts \cite{Nemec}.
Indeed, transport experiments carried out on metallic tubes with low
resistance contacts show ballistic transport and Fabry-P\'{e}rot
(FP)\ interference at low temperatures \cite{Liang2001}. In this regime,
coherence plays a major role and is manifested as oscillations in the
conductance as a function of the bias voltage. Besides the conductance
properties, the current noise has also been experimentally probed in this
regime \cite{Wu2007,Herrmann2007,YoungKim2007} as well as the effect of strong
magnetic fields \cite{Raquet2008}.

Motivated by these experiments, we decided to
consider the case of ac gating of a metallic nanotube-based Fabry-P\'{e}rot
resonator \cite{FoaTorres2009}. Our main question was whether control of the
current and noise could be achieved by tuning the parameters (such as
intensity and frequency) of a minimal ac field. To our surprise, our
subsequent research not only gave a precise and detailed answer to this
question but also pointed out a striking manifestation of the phase
sensitivity of the current noise \cite{FoaTorres2009}. We will give a closer
look to this topic in Sec. 3 but before we will briefly present our
theoretical tools.

\section{Tight-binding model and Floquet solution}

The theoretical approaches capable of describing time-dependent transport
include the Keldysh or non-equilibrium Green functions formalism
\cite{Pastawski1992,Jauho1994,Stafford1996,Arrachea2005b}, schemes that use
density functional theory \cite{Kurth2005}, the equation of motion method
\cite{Agarwal2007a}, and schemes that exploit the time-periodicity of the
Hamiltonian through Floquet theory
\cite{Moskalets2002,Camalet2003,Arrachea2006a}.

The main advantage in the Keldysh formalism is that interactions
(electron-electron and electron-phonon for example) can be included more
easily than in the other schemes. On the other hand, in the Floquet approaches
the picture is that of a single particle but the time-periodicity of the
Hamiltonian is fully exploited through the use of Floquet theorem thereby
simplifying the problem.

In our current research as a general framework we use the Floquet scheme
\cite{Camalet2003} combined with the use of Floquet-Green functions
\cite{FoaTorres2005}. Within this formalism, the dc component of the time
dependent current $I(t)$ can be computed as:
\begin{equation}
\bar{I}=\frac{2e}{h}\sum_{n}\int\left[  T_{R\longleftarrow L}^{(n)}%
(\varepsilon)f_{L}(\varepsilon)-T_{L\leftarrow R}^{(n)}(\varepsilon
)f_{R}(\varepsilon)\right]  d\varepsilon,
\end{equation}
where the transmission probabilities from left (L) to right (R) involving the
emission (absorption) of $n$ photons, $T_{R\longleftarrow L}^{(n)}%
(\varepsilon),$ can be fully written in terms of the Green functions for the
system \cite{Kohler2005,FoaTorres2005}. The current noise can be obtained from
the correlation function $S(t,t^{^{\prime}})=\left\langle \left[  \Delta
I(t)\Delta I(t^{\prime})+\Delta I(t^{\prime})\Delta I(t)\right]  \right\rangle
,$ $\Delta I(t)=I(t)-\left\langle I(t)\right\rangle $ being the current
fluctuation operator. The noise strength can be characterized by the zero
frequency component of this correlation function averaged over a driving
period, $\bar{S},$ which can be casted in a convenient way within this
formalism \cite{Kohler2005}. Further simplifications can be achieved by using
the broad-band approximation and an homogenous gating of the tube
\cite{Kohler2005}.

For simplicity we consider an infinite CNT described through a standard $\pi
$-orbitals Hamiltonian \cite{Charlier2007}%

\begin{equation}
H_{e}=\sum_{i}E_{i}^{{}}c_{i}^{\dagger}c_{i}^{{}}-\sum_{\left\langle
i,j\right\rangle }[\gamma_{i,j}c_{i}^{\dagger}c_{j}^{{}}+\mathrm{H.c.}],
\end{equation}
where $c_{i}^{\dagger}$ and $c_{i}^{{}}$ are the creation and annihilation
operators for electrons at site $i$, $E_{i}$ are the site energies and
$\gamma_{i,j}$ are nearest- neighbors carbon-carbon hoppings. The modelling of
the Fabry-P\'{e}rot interferometer is done by connecting a central part of length
$L$ (the \textquotedblleft sample\textquotedblright) to the rest of the tube
through matrix elements $\gamma_{t}$ smaller than the hoppings in the rest of
the tube which are taken to be equal to $\gamma_{0}=2.7$~$\mathrm{eV}$. \ In
the vicinity of the charge neutrality point, the dispersion relation is linear
and the mean level spacing of the isolated sample scales as $\Delta\propto
1/L$. A uniform ac gating of the sample can be modeled as an additional on
site energy $E_{j\in\mathrm{CNT}}=eV_{g}+eV_{\mathrm{ac}}\cos(\Omega t)$. In
the following section we will comment on some results obtained using this
general framework.

\section{Control of the conductance and noise in driven Fabry-P\'{e}rot
devices}

\begin{figure}[ptb]
\begin{center}
\includegraphics[width=15cm]{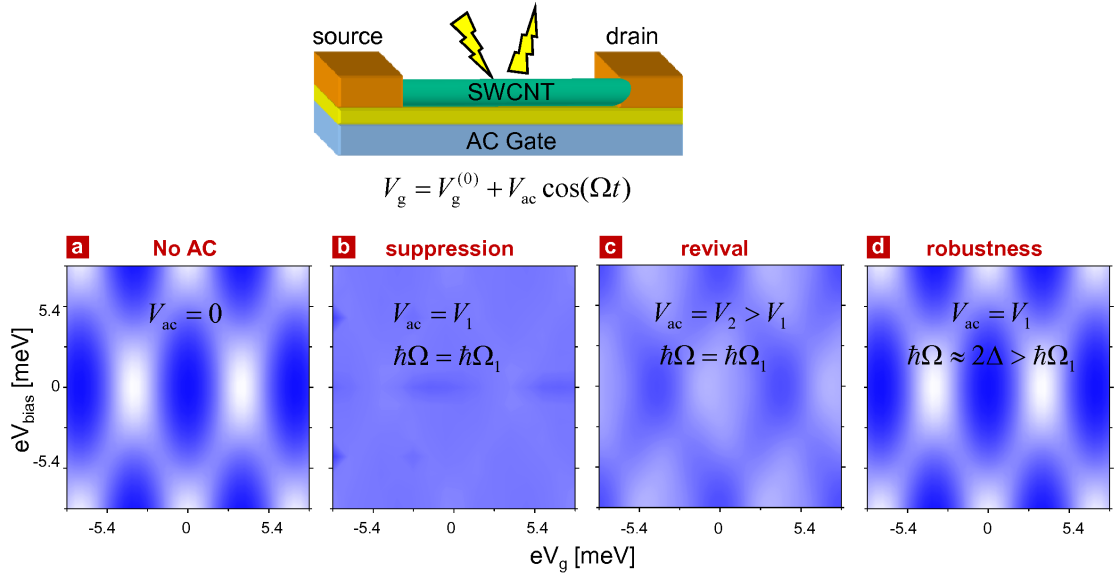}
\end{center}
\caption{(color online) Top: scheme of the system considered in the text, namely a SWCNT connected to source and drain electrodes and in the presence of an ac gating. Bottom: a-d are the Fabry-P\'{e}rot conductance patterns calculated for different values of the ac field frequency and amplitude (blue and white correspond to low and high conductances respectively).} \label{fig1} \end{figure}

Recently, we showed that the interplay between the ac field parameters (field
intensity and frequency) and the typical energy scales of the nanotube (such
as level spacing and position of van Hove singularities) can lead to strong
modifications of the conductance and current noise \cite{FoaTorres2009}. As
compared to the conductance interference pattern observed in static
conditions, by tuning the field intensity and frequency it can either be
suppressed, exhibit a revival, or show an ac-intensity independent behavior.
The former situation occurs whenever the ac field intensity attains specific
values that maximize the smoothing of the interference pattern introduced by
the inelastic processes, while the latter takes place whenever the driving
frequency is commensurate with the mean level spacing $\Delta$. This is
illustrated by the condutance interference patterns shown in Fig.
\ref{fig1}a-d. These patterns correspond to a numerical calculation for a 2 nm
diameter nanotube of $0.44~\mathrm{\mu m}$ length. The color scale (from dark
to light) is in the conductance range $(0.64,1)$ in units of the quantum limit
for the conductance of such a system, i.e., $4e^{2}/h$.

\begin{figure}[ptb]
\begin{center}
\includegraphics[width=15cm]{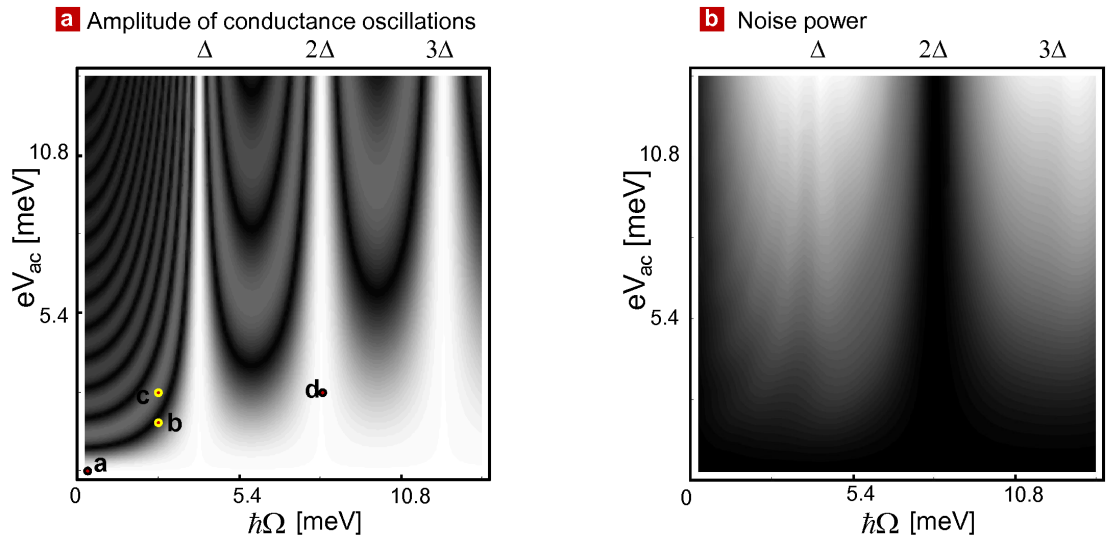}
\end{center}
\caption{(color online) a) Contour plot showing the amplitude of the
Fabry-P\'{e}rot oscillations as a function of the driving amplitude and frequency.
White stands for maximum amplitude and black for vanishing amplitude. b) Same
contour plot for the current noise $\bar{S}$, black is for vanishing noise.}%
\label{fig2}%
\end{figure}

The overall behavior of the conductance as we move in the driving
amplitude-frequency space is captured in \ref{fig2}a, where the half amplitude
of the conductance oscillations is shown in a color scale (white corresponds
to maximum amplitude and black to vanishing amplitude). There, the vertical
light regions indicate the regions where the interference pattern is
unaffected by the applied ac field. For low and intermediate frequencies, we
observe an alternating suppression and revival of the interference.

The situation in Fig. \ref{fig1}-d and \ref{fig2}a reminds us of the
wagon-wheel, stroboscopic or aliasing effect. The results of Fig. \ref{fig1}
hint that something similar occurs with the conductance patterns for our
Fabry-P\'{e}rot device, although here the energies or time-scales involved
belong to the quantum domain.

Based on the previous picture one would be tempted to think that whenever the
wagon-wheel condition is met both the current \textit{and }the noise should
behave as in the static case. However, numerical and analytical calculations
\cite{FoaTorres2009} of the current noise show that this is not the case.
Interestingly, the current noise in the low temperature and low bias limit
(which is purely due to the ac field) is suppressed whenever the frequency is
commensurate with \textit{twice} the mean level spacing as can be appreciated
in the contour plot of the current noise as a function of the ac amplitude and
frequency Fig. \ref{fig2}b. This happens due to the fact that the noise under
ac conditions is sensitive to the phase of the transmission amplitude, which
changes only by $\pi$ (and not by $2\pi$) from one resonance to the next one.
As emphasized in \cite{FoaTorres2009}, this constitutes a striking
manifestation of the phase sensitivity of the noise under ac conditions
leading to what we call the \textit{quantum wagon-wheel effect}.

Although the frequencies required to achieve the wagon-wheel condition are
here quite high (of about $100$ GHz), they are still in the order of what can
be experimentally achieved \cite{vanderWiel1999a} and can be reduced by
further tuning of the tube length and temperature. Besides, we must emphasize
that the main features on the left side of these amplitude-frequency maps (low
frequencies) persist even in the adiabatic limit. We point out that a similar
noise suppresion was reported in \cite{Moskalets2008} for a different system,
namely two barriers of varying strength and a uniform varying potential in
between. Noise suppression with maximum current was also discussed for a
different system in Ref. \cite{Strass2005}.

Again, we would like to stress that although we considered only the case of
carbon nanotubes our results remain valid for materials where the dispersion
relation may deviate from linear as discussed in the appendix and elsewhere
\cite{FTCGdR2009}.

\section{Conclusions}

An overview of some of the challenges and perspectives in the field of driven quantum transport in carbon-based devices was presented. The study of driven quantum transport in nanoscale devices is surely necessary (after all if these devices are going to be integrated in everyday electronics they will probably have to work under ac conditions). Moreover, ac fields also allow for a wealth of novel phenomena which can help to achieve control of the conductance, current
noise and may be even of the energy dissipation.

A particular example for the case of driven nanotube based Fabry-P\'{e}rot
devices was highlighted. We have shown that by tailoring the parameters of an
ac gating, important modifications can be achieved in both the conductance and
the current noise in all the frequency range. For frequencies commensurate
with twice the mean level spacing, both the conductance and the noise behave
as in the static case in a kind of \textit{quantum wagon-wheel effect}.

On the other hand, the phase sensitivity of the current noise
under ac conditions may offer interesting perspectives for the quantification
of the decoherence time \cite{Pastawski2002,FoaTorres2006a} in these low dimensional systems. This would require the modelling of decoherent processes.\medskip

\textit{Acknowledgements.} We acknowledge S. Kohler and M. Moskalets for
useful comments. This work was supported by the Alexander von Humboldt
Foundation, by the European Union project \textquotedblleft Carbon nanotube
devices at the quantum limit\textquotedblright\ (CARDEQ) under contract No.
IST-021285-2, and by the WCU (World Class University) program through the
Korea Science and Engineering Foundation funded by the Ministry of Education,
Science and Technology under contract R31-2008-000-10100-0. Computing time
provided by the ZIH at the Dresden University of Technology is also acknowledged.

\section{Appendix: the influence of non-linear dispersion}

In this appendix we address the issue of the effect of nonlinearities in the
energy dispersion in the phenomena reported in Sec. 3. A first observation is
that Fermi energies very close to the precise point where the dispersion
relation has a vanishing first derivative are not of interest because the
system does not behave as a proper metal and further gating is needed to observe
Fabry-P\'{e}rot interference. That said, we concentrate on metallic systems
with a generic dispersion relation close to the Fermi energy.

\begin{figure}[h]
\begin{center}
\includegraphics[width=15cm]{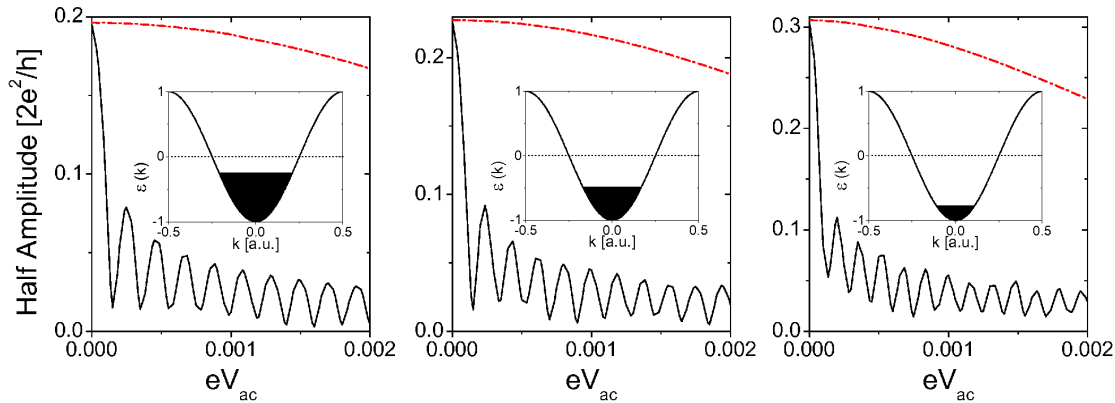}
\end{center}
\caption{(color online) Half amplitude of the Fabry-P\'{e}rot conductance
oscillations as a function of the amplitude of the ac-gating computed for a 1d
chain connected to electrodes. The panels are for different Fermi energies:
$\varepsilon=-0.25$ (left), $\varepsilon=-0.5$ (center) and $\varepsilon
=-0.75$ (right). The solid lines are for $\hbar\Omega=1.8\times10^{-4}$ and
the dash-dotted lines correspond to frequencies equal to the mean level
spacing. All the energies are given in units of half the band width (W). The
position of the Fermi energies are schematized in the insets. The change in
the vertical scale of the different panels is because at fixed coupling to the
leads, the escape rates to the leads change with energy, thereby modifying the
amplitude of the Fabry-P\'{e}rot oscillations.}%
\label{fig3}%
\end{figure}

For the case of carbon nanotubes, the dispersion relation at the charge
neutrality point is linear. As we move away from this point, the dispersion
relation acquires a non-linear component which naturally introduces dispersion
in the level spacings (as the levels of the finite nanotube are not all
equally spaced in the energy range of interest). This dispersion introduces
corrections to the simple picture presented in the manuscript to understand
the numerical results.

Indeed, a closer scrutiny at our numerical results \cite{FoaTorres2009} (which
were obtained for a full $\pi$-orbitals model for the nanotube) reveals that
the dips in the oscillations of the half amplitude of the FP conductance are
not perfectly zero as it would be for a system with constant level spacing.
The half amplitude of the FP oscillations at a frequency commensurate with the
level spacing is not perfectly constant neither.

When the relevant energies are further away from the perfectly linear
dispersion region, our simple picture will still be a good approximation
provided that the energy range effectively probed by the field ($\sim
\max(eV_{ac},N\hbar\Omega)$, where $N$ is the typical number of photons
excited by the field) is small enough. This can be realized by making a Taylor
expansion of the energy dispersion around the Fermi wave-vector:%
\[
\varepsilon(k)=\varepsilon_{F}+\left(  \frac{\partial\varepsilon}{\partial
k}\right)  _{k=k_{F}}(k-k_{F})+\left(  \frac{\partial^{2}\varepsilon}{\partial
k^{2}}\right)  _{k=k_{F}}(k-k_{F})^{2}+...,
\]
and requiring for the quadratic term to be smaller than the linear term. This
is always fulfilled in the low frequency/low ac-amplitude limit provided that
$(\partial\varepsilon/\partial k)(k=k_{F})\neq0$. In more formal terms,
$\max(eV_{ac},N\hbar\Omega)\ll\varepsilon_{F}$ is a sufficient requirement.

To illustrate this point we show in Fig. \ref{fig3} the results of
calculations for an ac-gated 1d system at different Fermi energies. Solid
black curves correspond to a frequency smaller than the mean level spacing
whereas the dash-dotted red lines are for a frequency equal to the level
spacing. As can be appreciated, the effect is quite robust even for Fermi
energies very close to the minimum of the dispersion relation.

\label{}






\end{document}